\documentclass[pra,aps,showpacs,twocolumn,amsmath,amssymb,floatfix]{revtex4}
\usepackage{epsfig}
\usepackage{graphicx}
\usepackage{amssymb,amsmath}
\usepackage{bbm}
\usepackage[dvips]{color}

\newcommand {\be}{\begin{equation}}
\newcommand {\ee}{\end{equation}}
\newcommand {\ba}{\begin{eqnarray}}
\newcommand {\ea}{\end{eqnarray}}

\begin{document}

\title{Spatial Dependence of Entropy in Trapped Ultracold Bose Gases}
\author{Lincoln D. Carr$^{1,2}$ and Markus K. Oberthaler$^{2}$}
\affiliation{
$^1$Department of Physics, Colorado School of Mines, Golden, CO 80401, USA \\
$^2$Kirchhoff Institut f\"ur Physik, Universit\"at Heidelberg, Im Neuenheimer Feld 227, 69120 Heidelberg, Germany}
\date{\today}

\begin{abstract}
We find a new physical regime in the trapped Bose-Hubbard Hamiltonian using time-evolving block decimation.  Between Mott-insulating and superfluid phases, the latter induced by trap compression, a spatially self-organized state appears in which non-local entropy signals entanglement between spatially distant superfluid shells.  We suggest a linear rather than harmonic potential as an ideal way to observe such a self-organized system.  We also explore both quantum information and thermal entropies in the superfluid regime, finding that while the former follows the density closely the latter can be strongly manipulated with the mean field.
\end{abstract}


\maketitle

Entropy is a fundamental concept in physics, chemistry, and information theory.  It is as foundational as energy for many-body systems, whether classical or quantum.  For example, consideration of entropy leads to the second law of thermodynamics, which is necessary to understand the efficiency of engines; thus entropy has had practical applications from the time of its invention in the nineteenth century.  In ultracold quantum gases, an excellent playground for quantum many-body physics and therefore basic many-body concepts such as entropy, the relationship between entropy and temperature for Bose and Fermi statistics is central to the concept of adiabatic cooling~\cite{carr2004d}; the latter led to the discovery of the atomic Fermi superfluid and profound new insight into how fermions pair to become bosons~\cite{bloch2008,regal2004,bartenstein2004}.  Moreover, entropy serves as an entanglement measure in such quantum systems~\cite{nielsenMA2000}.  However, in all these cases entropy is treated as a scalar -- a bulk measure.  In this Letter, we show that taking entropy as spatially dependent, whether as a zero temperature quantum information (QI) concept or a finite-temperature statistical mechanical concept, leads to new applications and new understanding of Bose gases.

%
\begin{figure}[t]
\begin{center}
\epsfxsize=8cm \epsfysize=6.5cm \leavevmode \epsfbox{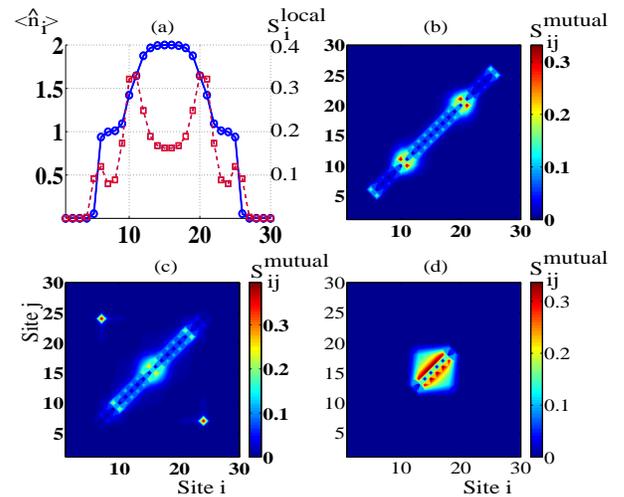}
\caption{\label{fig:weddingCake} (color online) \emph{Mutual entropy in the ``wedding cake'' regime of harmonic trap, where Mott-insulating layers are separated by superfluid shells}.  (a) The local entropy (dashed red curve) is oscillatory but non-zero in the MI layers; the density is shown for comparison (solid blue curve).  (b) The mutual entropy is oscillatory along the diagonal, indicating mainly nearest-neighbor entropy; the two bright spots are SF regions.  (c) When the system is compressed by the trap, highly nonlocal entropy appears.  (d) Further compression can lead to nonlocal entropy being dominant and completely SF system.}
\end{center}
\end{figure}

Specifically, we start with the trapped Bose-Hubbard Hamiltonian (TBHH)~\cite{fisher1989} at zero temperature.  Using time-evolving block decimation (TEBD), we explore three varieties of QI entropy: local von Neumann entropy, also called entropy of entanglement; block entropy; and mutual entropy, which is non-local.  The TBHH has two axes in the thermodynamic limit, namely chemical potential and tunneling, both in ratio to interaction.  Consideration of the local entropy shows that a third axis, trapping energy in ratio to interaction, is necessary to understand the TBHH.  We find a spatially self-organized state in this fundamentally finite system.  The characteristic feature of this new physical regime intermediate between superfluid (SF) and Mott Insulator (MI) is high non-local entropy between SF shells, as shown in Fig.~\ref{fig:weddingCake}(c).  Although this effect appears already for the harmonic potential, we suggest the linear potential as an ideal place in which to explore this idea.  Then we go on to explore the SF regime.  For a Thomas-Fermi (TF) profile~\cite{dalfovo1999} in the lattice plus harmonic trap, all three QI entropies closely follow the density.  On the other hand, thermal entropy behaves completely differently.  In the local density approximation (LDA) the entropy falls away rapidly in the region outside the condensate, but penetrates deeply into the condensate, having a small attrition of typically only about 50\% in the center; thus  a Feshbach resonance can be used to strengthen the mean field and push the entropy out to the edges of the condensate, thereby maximizing the efficiency of evaporative cooling below the critical temperature.

The first part of our study explores the TBHH outside the LDA.  Quantum phase transitions in finite systems where the LDA fails is key not only to fundamental features of quantum gases such as superfluidity~\cite{carr2009f} but also in nuclear physics~\cite{caprio2008}.
For the latter part of our study concerning thermal entropy in the LDA, a very recent work by Bernier \textit{et al.}~\cite{bernier2009} also proposes spatial manipulation of entropy, although purely in the LDA and for fermions.
They predict that an order of magnitude temperature decrease can be achieved, their main tool being dynamical mean field theory in the LDA.
For bosons, the study of thermal entropy as a bulk quantity has a long history and is well summarized in Dalfovo's review~\cite{dalfovo1999}.

We begin with a brief discussion of QI entropies.  The direct quantum analog of the thermal, or Gibbs entropy is the von Neumann entropy, also called the entropy of entanglement: $S_{\mathrm{vN}} = -\mathrm{Tr} \hat{\rho} \log \hat{\rho}$, where $\hat{\rho}$ is the state matrix, and indeed, for finite temperature we know that $S_{\mathrm{vN}} \to S_{\mathrm{Gibbs}}$ in the thermodynamic limit.  However, at zero temperature the von Neumann entropy is zero since the system is in a pure state.  Instead, we can ask about the local entropy based on local measurements of the on-site density matrix $\hat{\rho}_i \equiv \mathrm{Tr}_{j\neq i} \hat{\rho}$ in a lattice model.  In this vein, we define three distinct QI entropies:
\ba
S^{\mathrm{local}}_i & \equiv & -\mathrm{Tr} \hat{\rho}_i \log \hat{\rho}_i\,,\label{eqn:entropy1} \\
S^{\mathrm{block}}_i & \equiv & -\mathrm{Tr} \hat{\rho}_{<i} \log \hat{\rho}_{<i}\,,\\
S^{\mathrm{mutual}}_{ij} & \equiv & S_{ij} - S^{\mathrm{local}}_i - S^{\mathrm{local}}_j\,,
\label{eqn:entropy3} \ea
where $S_{ij}$ means tracing over all sites but two, and $\hat{\rho}_{<i}$ means tracing over all sites to the right of site $i$.  The local entropy is the nearest QI equivalent to the thermal spatial entropy density $S^{\mathrm{th}}(x)$.  The block entropy results from a bipartite splitting of the lattice made at site $i$.  The mutual entropy or mutual information~\cite{nielsenMA2000} assumes a measurement made at two sites only.

We now proceed to consider a harmonically trapped Bose gas in a lattice.  The TBHH, which assumes the tight-binding lowest-band approximation, is given by
\ba
\hat{H}&=& -J\textstyle\sum_{\langle i,j\rangle}(\hat{b}^{\dagger}_i \hat{b}_j+
\hat{b}_i \hat{b}^{\dagger}_j) +\textstyle\frac{1}{2}U\sum_i \hat{n}_i(\hat{n}_i-1)\nonumber\\
&&+\textstyle V_0 \sum_i i^p \hat{n}_i\,,
\hat{n}_i\,,\label{eqn:hubbard}
\ea
where $J$ is the hopping strength, $U$ is the interaction strength, $V_0 i^p$ is a power law potential with $p\in|\mathbb{Z}|$, and $\hat{b}_i$, $\hat{b}_i^{\dagger}$ are bosonic destruction/creation operators satisfying bosonic commutation relations on-site and commuting between sites.  We use TEBD~\cite{tebdOpenSource} in 1D to study the three QI entropies of Eqs.~(\ref{eqn:entropy1})-(\ref{eqn:entropy3}).  TEBD has two convergence parameters, the on-site dimension $d$, and the number of eigenvalues $\chi$ retained in the reduced density matrix.  In all cases we have checked convergence by at least doubling both $d$ and $\chi$ and seeing if we obtain the same results.  Our values of $d$ range from 8 to 12, meaning 7 to 11 bosons per site (8 to 12 including the vacuum), our values of $\chi$ range from 30 to 120, our number of sites is from $L=30$ to $100$, we use a number-conserving version of TEBD to make these large (d,$\chi$)-values tractable, and we implement imaginary time propagation in the Suzuki-Trotter scheme to obtain the ground state.

We first consider the MI regime, in which it has been shown that the harmonic trap causes the system to spatially traverse the TBHH phase diagram in the LDA, leading to a layered structure with SF shells separating MI layers~\cite{jaksch1998,batrouniGG2002}; this has been dubbed ``the wedding cake.''  Thermal entropy has been shown to be concentrated in the SF shells in this case, since the MI layers are a near-product state, and therefore have very small entropy; such considerations have led to new cooling schemes~\cite{hoTL2007}.  In contrast, for QI entropy, if the MI layer is less than about 15 sites, the entropy is much larger even far into the Mott lobe (more correctly, a ``Mott claw'' in 1D), due to finite size effects~\cite{carr2009i}; in this region, which is quite typical in experiments, the assumption of the LDA is of limited value.  However, we still find that the local entropy is large in SF shells and small in MI layers, varying by a factor of between 2 and 3 for $J/U=1/16$, $V_0/U=0.0125$, and $N=L=30$ bosons, for example, as shown in Fig.~\ref{fig:weddingCake}(a).  This is qualitatively similar to the thermal entropy result.  What about nonlocal entropy?  In Fig.~\ref{fig:weddingCake}(b) we show the mutual entropy corresponding to the parameters of Fig.~\ref{fig:weddingCake}(a).  The wedding cake structure is apparent: the mutual entropy is mainly nearest-neighbor and small in the MI layers, while it is larger and slightly more delocalized in the SF shells.

We have so far taken the harmonic trap to be sufficiently weak so that, even if not in the LDA, the MI layers are still many sites long.  What happens when we compress the trap?  This compression requires a third axis in the TBHH phase diagram and is a natural direction to explore in experiments.  For sufficiently strong traps the incompressibility of the MI phase is overcome and the system melts into a SF.  Considering such compression as a continuous process we observe in simulations that the MI layers melt gradually, shortening from many sites to just a few.  When each MI layer is reduced 2 or 3 sites we find a completely different form for the mutual entropy, as shown in Fig.~\ref{fig:weddingCake}(c) ($V_0/U=0.025$) and (d) ($V_0/U=0.5$).  The entropy becomes highly nonlocal due to penetration of SF tails through intervening MI layers.

\begin{figure}[t]
\begin{center}
\epsfxsize=8cm \epsfysize=8cm \leavevmode \epsfbox{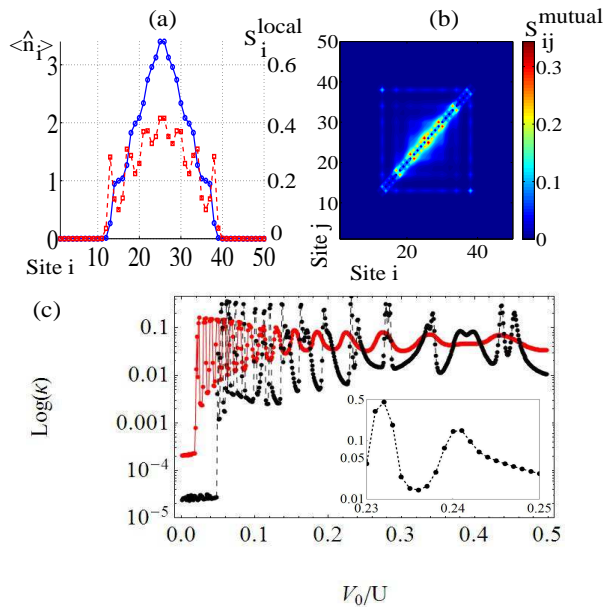}
\caption{\label{fig:linearPotential} (color online) \emph{New physical regime of the Bose-Hubbard Hamiltonian for a linear potential}.  (a) Density and local entropy, and (b) mutual entropy -- compare to the harmonic potential explored in Fig.~\ref{fig:weddingCake}.  (c) Compressibility $\kappa$ for $J/U=1/8$ (red points, solid curve) and $1/16$ (black points, dashed curve) as a function of the trap strength.  The inner panel shows a zoom around the value of $V_0/U=0.24$ used for panels (a) and (b).}
\end{center}
\end{figure}

In the harmonic trap this phenomenon is common but not regular, because a small amount of SF shell is required between each MI layer; due to the potential difference between sites being non-uniform (harmonic) this does not always occur.  In contrast, in a \emph{linear} potential the alternation between SF shells and MI layers is more regular and generic, since special parameters are not required and just compressing the trap naturally leads to this spatially self-organized state.
We use the word ``self-organization'' in the sense that the system optimizes arrangement of narrow SF and MI layers.  A typical example is shown in Fig.~\ref{fig:linearPotential}(a)-(b), where seven SF regions produce a nested square-like pattern in the mutual entropy.  In general, when the MI layers are a few sites long then there is a strong non-local mutual entropy, indicating that the MI layers act as effective potential barriers for the SF regions.  Thus there are three physical regimes: (1) for a weak linear trap the system is a MI; (2) for a stronger potential the bosons form alternating SF shells and MI layers, with an average wavelength $\bar{\lambda}$ which decreases monotonically as $V_0$ increases; and (3) completely melted, or SF, for very strong trap compression.  In Fig.~\ref{fig:linearPotential}(c) we show the compressibility $\kappa\equiv -\sigma^{-3}\partial \sigma/\partial(V_0)$ where $\sigma$ is obtained by a Gaussian fit.  After the abrupt change between from pure MI to wedding cake structure, the onset of strong oscillations indicates regime (2); such oscillations become longer wavelength in $V_0/U$ and damp out in the purely SF regime (3).  The sudden jumps in $\kappa$ have been discussed previously in the context of changing interaction~\cite{sengupta2005}; here we observe the same kind of effect caused by compressing with the trap.  The jumps are due to new superfluid regions entering the system.  Smaller $J/U$ leads to a stiffer response and therefore more sudden changes in $\kappa$.  The zoom shows that these changes are in fact smooth, simply more abrupt in some regions.  The outer panel of Fig.~\ref{fig:linearPotential}(c) has a resolution of 500 points and the derivative for $\kappa$ is approximated to second order.

%

\begin{figure}[t]
\begin{center}
\epsfxsize=8cm \epsfysize=6.5cm \leavevmode \epsfbox{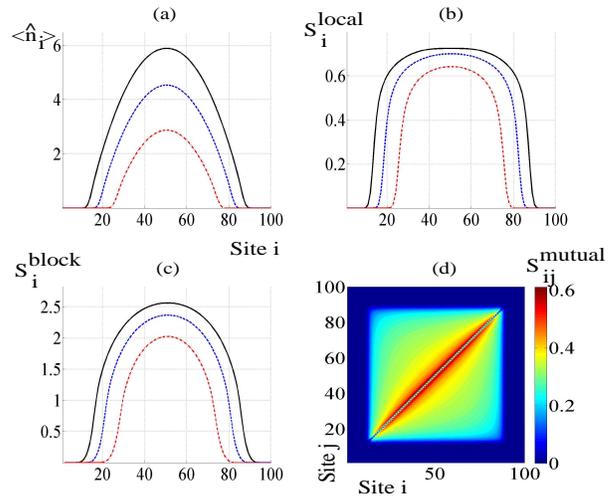}
\caption{\label{fig:superfluid} (color online) \emph{Quantum information entropies in the superfluid regime in a 1D lattice plus harmonic trap.}  Shown are (a) density, (b) local entropy, (c) block entropy, and (d) mutual entropy, all on 100 sites.  Panels (a)-(c) show $N=100,200,300$ atoms (red dash-dotted, blue dashed, and solid black curves); panel (d) shows 300 bosons only.}
\end{center}
\end{figure}

In order to make the comparison between QI entropy and the more familiar thermal entropy, we turn to the harmonic trap and the completely SF regime of Eq.~(\ref{eqn:hubbard}), $J/ZU \gtrsim 1$, where $Z$ is the coordination number.  We consider the TF regime in which the condensate takes a characteristic harmonic shape.  In this weakly discretized limit the lattice can provide a convenient discretization to approach the continuum problem in the finite trapped system by holding $U/\nu J =\gamma$ fixed, where $\nu$ is the filling and $\gamma$ is the Tonks parameter, and letting the number of sites be large~\cite{schmidt2007}.  We approach the SF continuum but remain still in the weakly discretized regime, with 100 sites and hundreds of bosons, where the LDA is meaningful.  Figure~\ref{fig:superfluid} shows the density and three QI entropies in this case, with $J=U=1$ and $V_0=0.004$.  The QI entropies closely follow the density -- note the TF profile, indicating that we are in the near-continuum limit.  Specifically, we find that the ratio of the full-width half-max (FWHM) of the entropies, local or non-local, to the FWHM of the density, is approximately constant over a wide range of parameters, studying $J/U$ from $1/2$ to 3 and a variety of system sizes and particle numbers -- the ratios of the FWHMs varies by only a few percent in any given case.

%

As we show in the following, the thermal entropy behaves completely differently from the QI entropy; it is even expelled from the center of the trap.  This is in contradistinction to
the MI regime, where both QI and thermal entropy showed the same behavior.
The \emph{spatial entropy} or local entropy density $S^{\mathrm{th}}(x)$ of a weakly interacting harmonically trapped Bose gas in the LDA in the semiclassical approximation and using the TF profile~\cite{dalfovo1999,carr2004d}, i.e., assuming $T\ll T_c$ and $\gamma\ll 1$ for our quasi-1D system~\cite{petrov2000b}, is given by
\be
S^{\mathrm{th}}(x) =  k_B\int\!\! \frac{dp}{2\pi\hbar} \left[
\beta \epsilon_\mathrm{b} \nu(p,x)
-\ln \left(1-e^{-\beta \epsilon_{\mathrm{b}}}\right)\right]\,,
\label{eqn:spatialEntropy}
\ee
where $\nu(p,x)=[\exp(\beta\epsilon_{\mathrm{b}})-1]^{-1}$ is the Bose weighting factor and
the energy of Boguliubov excitations is
\be \epsilon_{b}(p,x) \equiv
\left\{\!\!\sqrt{(p^2/2m)\left[p^2/2m+2g_{\mathrm{1D}}n(x)\right]}
\,,|x|\leq R \atop{p^2/2m+\frac{1}{2}m\omega^2 x^2
-\mu\,, \,\,\,\,\,\,\,\,\,\,\,\,\,\,\:\:\, |x|>R\, .}\right.
\label{eqn:bog}
\ee  In Eq.~(\ref{eqn:bog}), $\mu$ is the chemical potential, $R\equiv (2\mu/m\omega^2)^{1/2}$ is the TF radius, $n(x)$ is the mean-field density, $m$ is the atomic mass, $g_{\mathrm{1D}}\equiv \hbar \omega_{\perp} a_s$ is the contact interaction strength with $a_s$ the $s$-wave scattering length, and $\omega\ll \omega_{\perp}$, with $\omega_{\perp}$ the transverse trapping frequency.  The TF regime is reached for $\mu \gg \hbar \omega$, and involves neglecting the quantum pressure in the mean field theory, giving the condensate an inverted parabolic shape.

%
\begin{figure}[t]
\begin{center}
\epsfxsize=8cm \leavevmode \epsfbox{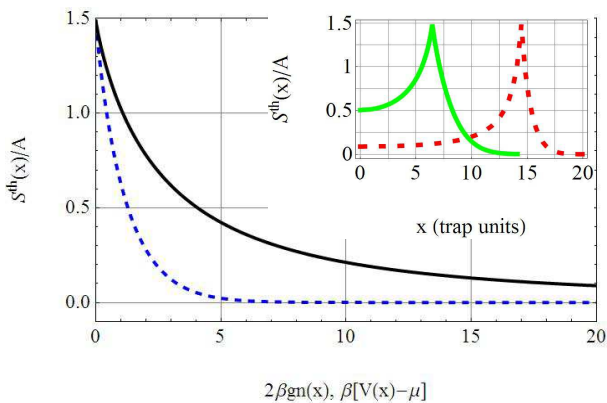}
\caption{\label{fig:thermal} (color online) \emph{Thermal entropy density for a BEC in a harmonic trap in the Thomas-Fermi regime}.  Outset: Spatial entropy density in the inner and outer regions of a Bose-Einstein condensate, as functions of $2\beta g n(x)\simeq 2\beta[\mu-V(x)]$ (solid black curve) and $\beta[V(x)-\mu]$ (dashed blue curve).  Note the rapid fall-off in the outer trap region compared to the slow fall-off in the inner, condensate region.  Inset: Distribution of the entropy (solid curves) for parameter range of Oberthaler experiments (green solid, dashed red curves).}
\end{center}
\end{figure}
In Fig.~\ref{fig:thermal} we show the numerical solution of Eq.~(\ref{eqn:spatialEntropy}) both inside and outside the TF radius, each integral being performed according to Eq.~(\ref{eqn:bog}), with the units of entropy per volume encapsulated in $A\equiv k_B (\sqrt{2}/\pi^2)(m/\beta\hbar^2)^{3/2}$.  Although the peak value of the entropy is always reached on the $y$-axis of Fig.~\ref{fig:thermal}, which occurs at the TF radius, the fall-off is not the same outside and inside the condensate.  In the outer region $S^{\mathrm{th}}(x)/A$ falls away very quickly as a function of $\beta [V(x)-\mu]$. In contrast, in the inner region it falls away much more slowly as a function of $2 \beta g n(x)$.  The extent of the fall-off is given by the peak condensate density as $2 \beta g n(0)=2\beta\mu$ in the TF regime.  For the experimental parameters of the Oberthaler group's $^{87}$Rb experiments, $\beta\mu$ ranges from 2 to 10, $\omega_{\perp}=2\pi\times 400$ Hz, $\omega=2\pi\times 20$ Hz, $a_s = 5.29 \times 10^{-7}$ cm, $k_B T \simeq 10$ to 20 nK, and there are 1000 to 2000 bosons.  Then $2 \beta  g n(0) =4$ to 20 and the entropy falls off by between 70\% and 93\% from the edge of the condensate to the center, as sketched in the inset of Fig.~\ref{fig:thermal} in ``trap units'' $\equiv \sqrt{\hbar/m\bar{\omega}}$.  A Feshbach resonance can be used to increase this range greatly, localizing the trap entropy even more strongly at the trap edges for more effective evaporative cooling below the critical temperature.  After an evaporative cooling step the interactions can be decreased to avoid strong losses incurred by the Feshbach resonance.  Adiabatic and cyclical transport of the entropy in this manner, in conjunction with a careful study of loss effects~\cite{carr2004e}, provides a subject for future theoretical and experimental investigations.

In past studies of the wedding cake structure, phase coherence between SF shells was neither expected nor observed.  In the new highly-compressed regime we suggest, an interference experiment performed by turning off the trap and letting the system expand should result in an observable periodic interference pattern on top of the smeared-out pattern corresponding to the MI.  In repeated experiments the position of the peaks should be stable due to phase coherence.  Recent 3D LDA studies of Josephson oscillations between SF layers separated by MI plateaus suggest that such phenomena are not restricted to 1D~\cite{vishveshwara2008,rigol2009}.

In conclusion, we explored spatial control of entropy in trapped Bose gases.  We used quantum information entropy to find a new axis of the trapped Bose-Hubbard Hamiltonian outside of the LDA, namely the harmonic trap strength.  Between MI and SF phases, the latter induced by trap compression, we found a new spatially self-organized state signaled by large non-local entropy; we suggested the linear potential as an ideal trap in which to observe this effect.  Turning to the weak-lattice superfluid regime where LDA is valid, we found that QI and thermal entropies do not display the same pattern.  Although QI entropy closely follows the boson density, thermal entropy can be localized at the trap edges, depending on the mean field strength; this leads to new cooling applications for the future.

We acknowledge use of the time-evolving block decimation algorithm from the TEBD open source project~\cite{tebdOpenSource} and thank Michael Wall.  This work was supported by the National Science Foundation under Grant PHY-0547845 as part of the NSF CAREER program, by the Graduate School of Fundamental Physics at the University of Heidelberg, by the Aspen Center for Physics, and by the ExtreMe Matter Institute EMMI in the framework of the Helmholtz Alliance HA216/EMMI.


\end{document}